*Article*

# Examination of Eye-Tracking, Head-Gaze, and Controller-Based Ray-casting in TMT-VR: Performance and Usability Across Adulthood

**Panagiotis Kourtesis[1-4]\*[†], Evgenia Giatzoglou[1][†], Panagiotis Vorias[1], Katerina Alkisti Gounari [1], Eleni Orfanidou[1], and Chrysanthi Nega[1]**

[1] Department of Psychology, The American College of Greece, 15342 Athens, Greece
[2] Department of Informatics & Telecommunications, National and Kapodistrian University of Athens, 16122, Athens, Greece
[3] Department of Psychology, National and Kapodistrian University of Athens, 157 84 Athens, Greece
[4] Department of Psychology, The University of Edinburgh, Edinburgh EH8 9Y, UK
\* Correspondence: pkourtesis@acg.edu
[†] These authors contributed equally to this work.

**Abstract:** Virtual reality (VR) can enrich neuropsychological testing, yet the ergonomic trade-offs of its input modes remain under-examined. Seventy-seven healthy volunteers—young (19–29 y) and middle-aged (35–56 y)—completed a VR Trail-Making Test with three pointing methods: eye-tracking, head-gaze, and a six-degree-of-freedom hand controller. Completion time, spatial accuracy, and error counts for the simple (Trail A) and alternating (Trail B) sequences were analysed in 3 × 2 × 2 mixed-model ANOVAs; post-trial scales captured usability (SUS), user experience (UEQ-S), and acceptability. Age dominated behaviour: younger adults were reliably faster, more precise, and less error-prone. Against this backdrop, input modality mattered. Eye-tracking yielded the best spatial accuracy and shortened Trail A time relative to manual control; head-gaze matched eye-tracking on Trail A speed and became the quickest, least error-prone option on Trail B. Controllers lagged on every metric. Subjective ratings were high across the board, with only a small usability dip in middle-aged low-gamers. Overall, gaze-based ray-casting clearly outperformed manual pointing, but optimal choice depended on task demands: eye-tracking maximised spatial precision, whereas head-gaze offered calibration-free enhanced speed and error-avoidance under heavier cognitive load. TMT-VR appears to be accurate, engaging, and ergonomically adaptable assessment, yet it requires age-specific–stratified norms.

## 1. Introduction

Cognitive instruments such as the Trail Making Test (TMT) remain staples for indexing task-switching, processing speed, and visual attention [1]. Yet both paper and 2-D computer formats fall short on ecological validity—the twin criteria of verisimilitude and veridicality—because they strip away the spatial depth, distractors, and time pressure that define everyday behaviour [2,3]. Consequently, TMT scores correlate only modestly with functional activities, leaving a gap between laboratory findings and real-world competence [4,5].

Immersive virtual reality (VR) offers a corrective. Modern head-mounted displays situate users in interactive, precisely logged 3-D scenes, preserving internal control while boosting ecological fidelity [6,7]. Integrated eye-tracking and sub-millisecond telemetry support granular reaction-time analytics, cloud storage standardises administration, and automated feedback removes rater error [8]. VR assessments can embed complex tasks—navigating crowded streets, multitasking under simulated deadlines, dodging unexpected obstacles—that are impractical in clinics yet critical for diagnostic specificity [3,9]. Gamified elements sustain engagement and reduce test anxiety, while parameter tuning enables both standardisation and individualisation [10,11].

The VR adaptation of the TMT (TMT-VR) illustrates these advantages. Studies show it delivers higher data accuracy, millisecond-timed performance metrics, and markedly stronger ecological validity than the legacy test [12]. Beyond executive speed, immersive versions have begun probing episodic memory and sustained attention, broadening the cognitive domains accessible in realistic settings [13]. By aligning assessment conditions with everyday demands, VR tools promise more sensitive diagnostics, earlier interventions, and wider accessibility through remote or self-administered formats [3,14,15]. In short, immersive VR is not merely a technological novelty but a pragmatic evolution that can finally bridge psychometric rigour and real-world relevance.

*Time & Accuracy in TMT-VR*

The Trail Making Test remains a sensitive gauge of executive speed (TMT-A) and set-shifting (TMT-B) in normal ageing and early Alzheimer's disease [16,17]. In its immersive adaptation, TMT-VR retains these metrics—completion time and error rate—while adding 3-D depth and continuous logging, but performance must now be interpreted through two new lenses: interface demand and digital familiarity [3,18].

Time costs grow with age and fall with education in the paper test; a college degree can offset a decade of normative slowing [1,19]. Similar trends persist in VR, though the gap narrows for everyday-like tasks [14,20]. Starting in the fifth decade, VR completion times diverge by roughly 20 years of chronological age [21]; younger adults also outpace older ones on other immersive batteries [22–24]. Still, digital fluency modulates both desktop and VR scores—high-tech users finish faster and err less in Parts A and B [25].

Accuracy shows a different profile. Older adults often trade speed for safety, lifting pen or controller to avoid mistakes [26]. Always-on eye- or head-tracking in TMT-VR reduces such pauses, exposing genuine processing limits. We therefore follow VR norms: more errors imply weaker executive control, and longer times indicate slower processing [27]. Because executive tasks rely on working memory, inhibition, and cognitive flexibility [28,29], any modality that cuts motor–cognitive dual demands may blunt age effects [10,14]. Our study tests that premise directly, asking whether eye-tracking or head-gaze can level the playing field across age and education.

Additionally, evidence on whether habitual gaming boosts VR-based cognitive test scores is mixed. Some work links regular play—especially in middle adulthood—to sharper executive control and problem-solving [30,31]. Other studies, however, report no gamer edge on cognitive-motor tasks [26,32]. A recent TMT-VR validation likewise found no performance lift for young adult gamers versus non-gamers [33], challenging the assumption that controller proficiency automatically yields higher scores. Gamers do own headsets more often [34] and may navigate virtual spaces with less emotional load [35] or faster spatial learning [36], yet these benefits seem confined to controller-based input. Hands-free modalities such as eye-tracking or head-gaze largely equalise outcomes, underscoring the importance of interface design for equitable assessment [10,14].

*Interaction Modalities in TMT-VR: Precision vs Ergonomics*

Modern VR testing is shifting from hand-held controllers to hands-free inputs that better mirror natural behaviour and reduce motor load [11,14]. In cognitive batteries such as TMT-VR, three inputs dominate—eye-tracking, head-gaze, and 6 DoF controllers—each with distinct trade-offs that must be weighed against user age, digital fluency, and motor capacity [12,33].

Eye-tracking records gaze vectors via infrared corneal reflection, yielding millisecond precision with minimal physical effort [37,38]. Hands-free interaction modalities consistently outperform controller input on speed, accuracy, and/or cognitive workload in both general HCI tasks and TMT-VR [33,39]. Reliability, however, hinges on stable calibration and good pupil detection; reflections, ptosis, or poor lighting can degrade performance [40]. Despite these caveats, improving hardware makes gaze the front-runner for low-fatigue, ecologically valid assessment [41].

Head-gaze steers a reticle with head orientation and confirms via dwell [42]. It removes controllers but taxes cervical muscles and may slip on users with limited neck mobility [39]. In structured, small-field tasks (e.g., TMT-VR-B) head-gaze can match or exceed controller times, yet remains slower and/or less precise than gaze in dynamic scenes [33,43].

Controllers provide tactile feedback and familiar game metaphors [11,44] but introduce arm fatigue and add variance tied to gaming skill [18,45]. Usability studies cite frequent errors in button mapping and object release, making controllers the weakest option for purely cognitive metrics [46].

All modalities may facilitate ray-casting—a virtual laser pointer— which offers fast, distance-independent selection with clear visual feedback, especially valuable for older adults or users with tremor [41,47]. When calibrated, raycasting feels natural and curbs motor strain [48,49]. In sum, eye-tracking currently provides the best balance of accuracy and ergonomics; head-gaze is a viable, calibration-free fallback; and controllers should be reserved for training or as a last-resort input. Rigorous, modality-specific evaluation—especially in middle-aged users—remains essential for equitable VR neuropsychological assessment [14,50].

*Study Aims and Scope*

We evaluated a VR version of the Trail Making Test (TMT-VR) in healthy young (18-35 yr) and middle-aged (36-59 yr) adults, asking how three inputs—eye-tracking, head-gaze, and 6-DoF controllers—influence classic TMT metrics (completion time, errors). Parallel analyses assessed System Usability Scale, brief UX, and acceptability scores, then tested whether prior gaming skill altered either performance or perceptions. By targeting middle age—a cohort rarely studied but already showing executive slowing—we extend earlier TMT-VR work confined to students and confirm whether hands-free modalities genuinely ease cognitive load.

Research Questions

1. Do input modes differentially affect speed and accuracy across the two age groups?
2. Do usability, UX, and acceptability ratings differ by age?
3. Does gaming experience moderate performance or usability?

Hypotheses

- H1: Eye-tracking and/or head-gaze will surpass controller for both speed and accuracy.
- H2: Both age groups will rate usability high
- H3: Gaming experience will not materially alter task metrics or subjective ratings, reflecting interface inclusivity.

## 2. Materials and Methods

*Participants*

Recruitment was conducted at the Psychology Network Lab (PsyNet Lab), Department of Psychology, American College of Greece. Flyers, mailing-lists, and QR-code sign-ups invited volunteers who were (a) 18–34 years ("young") or 35–59 years ("middle-aged"), (b) fluent in English ("bilingual proficiency or native speakers"), and (c) free of neurological, psychiatric, visual, or motor disorders. An a-priori calculation for a 3 (modality) × 2 (age) mixed design with medium effect ($\eta^2_p = .06$), $\alpha = .05$, and $1-\beta = .80$ suggested n = 72. Eighty-two individuals enrolled; five were screened out due to eye-tracking calibration failure, leaving N = 77 (39 young, 38 middle-aged). Mean ages were 23.49 ± 2.66 yrs and 42.55 ± 4.95 yrs; 50.6 % identified as female. Gaming proficiency was quantified with the Gaming Skill Questionnaire (GSQ) [51,52]. Total scores ($\alpha \geq .70$) were median-split (Mdn = 22) into High-Gaming (n = 38) and Low-Gaming (n = 39) sub-samples balanced for age group and gender ($\chi^2 < 1$, ns).

Gaming Skill Questionnaire (GSQ)

To quantify prior video-gaming experience we administered the GSQ, a 36-item self-report inventory that surveys six popular genres—sports, first-person shooters, role-playing, action-adventure, strategy and puzzle titles—each with parallel frequency and skill items [51,52]. Participants judged how often they played specific videogame genres and how competent they felt when doing so, using six-point Likert anchors that ranged from "less than once per month / no experience" to "every day / expert." Genre scores were calculated as the product of the frequency and skill ratings, then summed to create a total gaming index (possible range 6–216). Internal consistency in the present sample was high (Cronbach's $\alpha = .88$), replicating the psychometric structure reported in the original validation. Because the subsequent analyses contrasted higher- vs lower-experience gamers, a median split of the total index served to balance cell sizes across the factorial design.

Cybersickness in VR Questionnaire (CSQ-VR)

To monitor simulator side-effects we used the CSQ-VR, a streamlined six-item instrument that asks respondents to rate nausea, vestibular disturbance and oculomotor strain on a seven-point intensity continuum [53]. Each subscale yields a score between 2 and 14, while the aggregated total ranges from 6 to 42, with larger values indicating more severe cybersickness. The scale was administered twice—immediately before donning the headset and again after removing it—to gauge any change attributable to exposure. Reliability coefficients in the current cohort echoed earlier reports (pre-exposure $\alpha = .86$; post-exposure $\alpha = .84$), and no participant exceeded the recommended threshold for moderate cybersickness (total > 18).

*Virtual-Reality Apparatus*

All testing was carried out in the PsyNet Lab using a hardware configuration explicitly recommended for minimizing cybersickness during prolonged immersive exposure [54]. A high-performance PC (Intel i9-13900KF, 32 GB RAM, NVIDIA RTX 3080) powered the virtual scene, ensuring a frame-rate that never fell below 90 Hz and thereby preventing the visual–vestibular delays known to provoke disorientation. Visual stimuli were delivered through a Varjo Aero head-mounted display whose dual mini-LED panels provide 2880 × 2720 pixels per eye at a 90 Hz refresh rate, coupled with custom aspheric lenses that suppress peripheral distortion. Two SteamVR Lighthouse 2.0 base stations, mounted diagonally at ceiling height, afforded sub-millimetre head- and controller-tracking precision across the entire two-metre test space. Participants wore over-ear, active-noise-cancelling headphones so that all auditory feedback originated exclusively from the software, further isolating them from external distractions. Interaction in the controller modality employed the HTC Vive 6-DOF wands; their balanced weight distribution, textured grips and analogue triggers have repeatedly been shown to support rapid and accurate pointing in room-scale tasks [55].

*Development and Ergonomic Optimisation of the TMT-VR*

Construction of the multimodal TMT-VR followed ISO 9241-210 guidelines for human-system ergonomics and adhered to published recommendations for cognitive testing in immersive environments [32,56]. A user-centred design framework directed every iteration: early prototypes were piloted by naïve volunteers who provided think-aloud feedback on visual clarity, dwell-time thresholds and controller comfort; resulting comments were recycled into subsequent Unity 2022.3 builds until no additional usability issues emerged.

Tasks A and B were rendered inside a deliberately feature-poor, neutral-grey chamber that minimised extraneous memory cues while accentuating the depth gradient afforded by three concentric shells (radii = 3, 6 and 9 m) and three elevation planes (0°, +15°, +30°). Twenty-five numbered or alphanumeric spheres (0.25 m diameter) spawned at pseudo-randomised polar coordinates, guaranteeing that no participant confronted an identical array. A single ray-casting routine controlled selection across modalities: in head- and eye-gaze modes the ray emanated from the headset's forward vector or the Varjo 120 Hz gaze point, respectively; in the controller mode it projected from the wand's tip. To avoid the Heisenberg effect—performance decrements produced by simultaneous pointing and button-press confirmation [57]—a hit was registered automatically after a continuous 2 s fixation. The same dwell logic mitigated the "Midas-touch" problem of inadvertent target activation, and the sphere edges were always left marginally visible so that subtle head movements could disambiguate partially occluded items [58].

Because colour-vision deficiency can undermine visual-search tasks, all feedback cues conformed to established colour-blind-safe palettes [59]. Correct selections turned a saturated yellow accompanied by a pleasant chime; erroneous ones flashed red with a short buzzer. Haptic feedback was intentionally disabled to prevent confounding motor cues, preserving comparability across the three interaction modes.

Tutorials were embedded as interactive scenes preceding each experimental block. Spoken instructions (studio-recorded at −23 LUFS) guided participants through sample arrays until they completed three consecutive selections without error, ensuring comprehension while eliminating experimenter-delivery variance. Eye calibration used Varjo Base's two-stage procedure; sessions exceeding a 1.2° root-mean-square error were recalibrated.

This integrated hardware–software ecosystem therefore satisfied three principal aims: (a) it delivered high-fidelity optics and low-latency tracking that curb cybersickness; (b) it implemented selection mechanics grounded in human-factor evidence to maximise ecological validity; and (c) it maintained methodological parity with the traditional TMT so that performance indices—completion time, spatial accuracy and error count—could be interpreted against an established neuropsychological benchmark.

Demographic and Technology-Use Form

Upon arrival, volunteers completed a short paper-and-pencil questionnaire that recorded age in years, sex, and highest educational attainment. Two continuous visual-analogue items (0 = "never" to 10 = "every day, expert level") captured habitual personal-computer and smartphone usage.

Trail Making Test – Virtual Reality (TMT-VR)

Participants completed two tasks that paralleled traditional TMT-A and TMT-B. In Task A they selected the integers 1-through-25 in ascending order; in Task B they alternated between numbers and letters (1-A-2-B … 13). Each task could be executed with three interaction modalities presented in counter-balanced order. In the eye-tracking mode, Varjo's 120 Hz gaze stream generated an invisible ray that triggered a selection when cumulative fixation on a target exceeded 1.5 s; blinks were filtered out using the manufacturer's pupil-diameter dropout flag. In the head-gaze condition, the camera forward-vector controlled a central reticle that employed identical dwell-time logic, ensuring the only difference between modalities was the effector itself. In the 6-degree-of-freedom controller mode, participants pointed a laser from an HTC-Vive wand. Across all modes a correct object turned yellow, whereas an erroneous hit flashed red and an auditory buzzer sounded, prompting the participant to resume from the last correct sphere.

Three performance indices were extracted automatically: total completion time expressed in milliseconds, mean spatial accuracy calculated as the Euclidean distance between ray-impact and sphere centre at the moment of selection, and an error count summarising out-of-sequence touches. To guard against practice effects, the spawn algorithm ensured that no two participants encountered the same spatial layout. All interaction tutorials were embedded inside the virtual scene via video screen and voiced via noise-cancelling headphones, a strategy that standardised instruction delivery and reduced experimenter bias. A video presentation of each task of TMT-VR can be accessed using the following links: TMT-VR Task A and TMT-VR Task B (accessed on 14/6/2024).

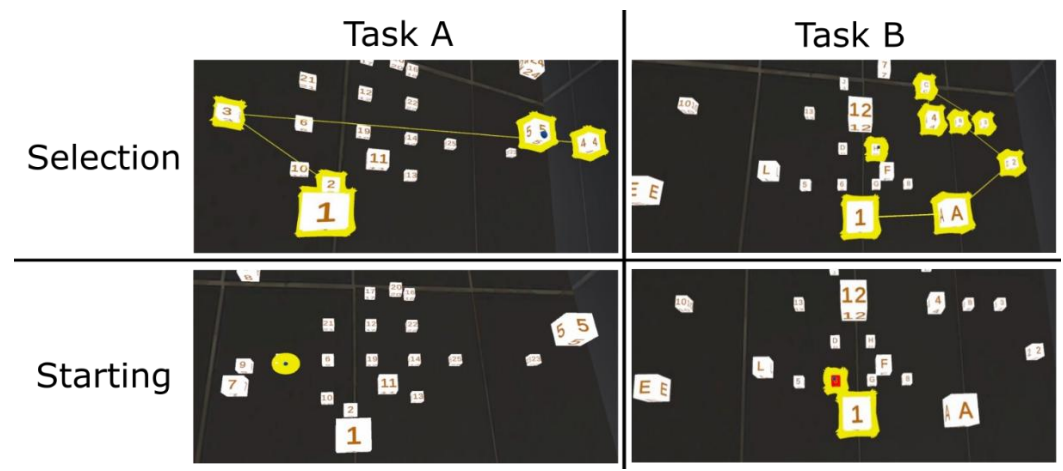

**Figure 1.** The Tasks of the TMT-VR. Task A (Left), Task B (Right), Starting Position (Bottom), and Selection of Targets (Top). Figure derived from Giatzoglou et al., (2024) [33]

Subjective Evaluation Scales of Usability, UX, and Acceptability

Immediately after completing the six experimental trials, participants removed the headset and filled out three paper questionnaires. Usability perceptions were captured with the System Usability Scale (SUS), a ten-item, five-point inventory whose robust factor structure and sensitivity to subtle interface flaws are well documented [60,61]. The short version of User Experience Questionnaire (UEQ-Short) provided a multidimensional evaluation of user experience, asking respondents to rate 26 semantic differential pairs such as "obstructive–supportive" or "conventional–innovative" on a seven-point continuum. In the present data set, internal reliability for the attractiveness meta-factor was .89, while pragmatic and hedonic facets ranged between .73 and .82, aligning with published norms [62]. Acceptability was gauged with the ten-item public-use adaptation of the Service User Technology Acceptability Questionnaire (SUTAQ), which has demonstrated convergent validity with adoption intentions in prior VR health applications [63,64]. Items such as "The VR test would fit easily into my daily life" were answered on a six-point agreement scale; Cronbach's alpha in this study was .82.

Procedure

The study followed the Declaration of Helsinki and received departmental ethics approval. The experimental protocol was implemented in the PsyNet Lab of the Department of Psychology, where all testing stations were configured identically to guarantee stimulus fidelity and consistent timing across sessions.

On arrival, participants signed informed consent, then completed demographics, PC/SMART use, GSQ, and baseline CSQ-VR. A two-stage Varjo Base calibration (static dot → dynamic dot) ensured gaze precision. Participants stood on a marked 3 m × 3 m VR area; experimenters stabilised cables and posture. Six tutorial blocks (Task A/B × three modalities) standardised instructions and minimised learning effects [11].

The main experiment comprised three blocks; each block used one modality and presented Task A then Task B. Block order was Latin-square counter-balanced by an automated scheduler to blind experimenters. Verbal reminders emphasised both speed and accuracy. Between blocks a 30-s "visual reset" scene mitigated fatigue and cybersickness. Total VR exposure never exceeded 20 min.

Immediately post-VR, participants re-completed CSQ-VR, then SUS, UEQ-S, and Acceptability forms in Qualtrics. A structured debrief allowed questions about the experiment. Average visit duration was 35 ± 5 min.

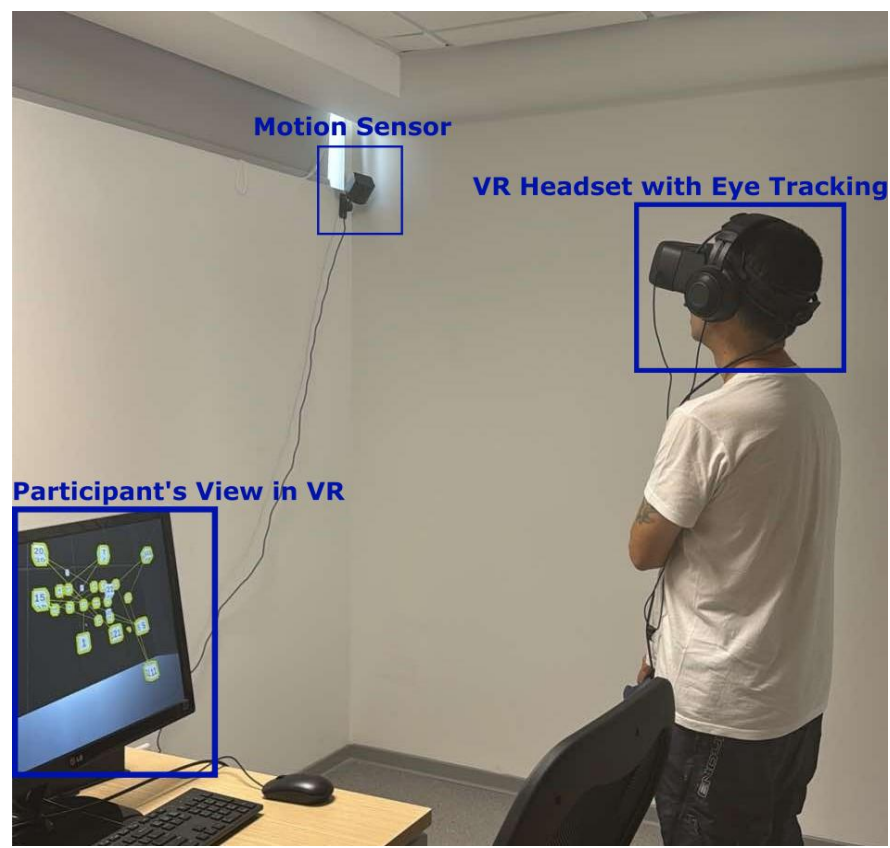

**Figure 2.** Participant Performing TMT-VR using Eye Tracking Based Interaction. Figure derived from Giatzoglou et al., (2024) [33]

Statistical Analysis

All analyses were carried out in R [65] via RStudio [66]. Variable distributions were inspected with the Shapiro–Wilk test, and any that deviated from normality were transformed with bestNormalize [67]. We then calculated means, standard deviations, and ranges for every demographic, performance, and questionnaire variable. Exploratory figures generated in ggplot2 [68] provided an overview of group trends and were later reused to visualise significant model effects.

Bivariate relations among age, education, gaming-skill scores, TMT-VR completion times, spatial accuracy, and subjective ratings (GSQ, SUS, UEQ-S, AQ) were quantified with Pearson correlations. Resulting matrices were displayed with corrplot [69] to facilitate interpretation.

Primary hypotheses were tested with two 3 × 2 mixed repeated-measures ANOVAs implemented in afex [70]. Interaction Mode (eye-tracking, head-gaze, controller) served as the within-subjects factor, while Gaming Skill (high vs. low, median split) was the between-subjects factor—analysed separately for Tasks A and B. When sphericity was violated, Greenhouse–Geisser corrections were applied. Post-hoc pairwise contrasts were obtained with the emmeans package [71], using Bonferroni adjustment to maintain the familywise error rate. All effect sizes are reported as partial $\eta^2$ (ANOVA) or Cohen's d (post-hoc comparisons), and statistical significance was set at $p < .05$.

## 3. Results

All the descriptive statistics are displayed in Table 1,2 and 3.

**Table 1.** Demographics per Age-Group & Gaming Level

| Variable | Age Group | Gaming Level | Mean (SD) | Range |
|---|---|---|---|---|
| Age | Middle | High | 40.90 (3.76) | 35–48 |
| | Young | High | 23.65 (2.43) | 19–29 |
| | Middle | Low | 44.21 (5.45) | 35–56 |
| | Young | Low | 23.32 (2.89) | 19–29 |
| Education | Middle | High | 16.90 (2.64) | 12–22 |
| | Young | High | 16.50 (1.08) | 14–18 |
| | Middle | Low | 17.16 (2.84) | 12–24 |
| | Young | Low | 16.63 (2.27) | 14–20 |
| PC | Middle | High | 10.37 (0.99) | 8–12 |
| | Young | High | 10.15 (0.92) | 8–12 |
| | Middle | Low | 9.95 (1.11) | 8–12 |
| | Young | Low | 9.26 (1.30) | 6–11 |
| SMART | Middle | High | 10.21 (1.21) | 8–12 |
| | Young | High | 9.85 (1.47) | 6–11 |
| | Middle | Low | 9.84 (1.05) | 8–12 |
| | Young | Low | 10.05 (1.41) | 7–12 |
| VR | Middle | High | 3.58 (1.55) | 2–7 |
| | Young | High | 2.35 (0.58) | 2–4 |
| | Middle | Low | 2.42 (0.68) | 2–4 |
| | Young | Low | 2.26 (0.44) | 2–3 |

PC = Computing Experience; SMART = Smartphone Application Use Experience; VR = VR Use Experience

**Table 2.** Genre-Specific Gaming Experience per Age-Group & Gaming Level

| Variable | Age Group | Gaming Level | Mean (SD) | Range |
|---|---|---|---|---|
| SPORT | Middle | High | 6.16 (2.10) | 2–10 |
| | Young | High | 4.90 (2.14) | 2–10 |

|  | Middle | Low | 2.68 (0.93) | 2–5 |
|---|---|---|---|---|
|  | Young | Low | 2.32 (0.47) | 2–3 |
| FPS | Middle | High | 4.90 (2.74) | 2–10 |
|  | Young | High | 5.55 (2.31) | 3–10 |
|  | Middle | Low | 2.42 (1.24) | 2–6 |
|  | Young | Low | 2.37 (0.49) | 2–3 |
| RPG | Middle | High | 4.16 (2.86) | 2–11 |
|  | Young | High | 5.05 (2.64) | 2–10 |
|  | Middle | Low | 2.00 (0.00) | 2–2 |
|  | Young | Low | 2.00 (0.00) | 2–2 |
| Action | Middle | High | 6.05 (2.50) | 2–10 |
|  | Young | High | 4.45 (1.90) | 2–8 |
|  | Middle | Low | 2.00 (0.00) | 2–2 |
|  | Young | Low | 2.53 (0.89) | 2–5 |
| Strategy | Middle | High | 4.05 (1.11) | 2–6 |
|  | Young | High | 3.45 (1.76) | 2–8 |
|  | Middle | Low | 2.00 (0.00) | 2–2 |
|  | Young | Low | 2.00 (0.00) | 2–2 |
| Puzzle | Middle | High | 5.37 (3.04) | 2–12 |
|  | Young | High | 3.70 (1.69) | 2–7 |
|  | Middle | Low | 2.63 (1.28) | 2–6 |
|  | Young | Low | 2.21 (0.41) | 2–3 |

SPORT =Sports Games ; FPS = First Person Shooting Games; RPG = Role Playing Games; Action = Action-Adventure Games; Strategy =Strategy Games; Puzzle =Puzzle Games

**Table 3**. Sex Distribution by Age Group

| **Sex** | **Age Group** | **n** | **%** |
|---|---|---|---|
| Female | Middle | 21 | 27.3 % |
|  | Young | 23 | 29.9 % |
| Male | Middle | 17 | 22.1 % |
|  | Young | 16 | 20.8 % |

Percentages are calculated relative to the total sample (*N* = 77).

*Correlational Analyses*

Table 4 summarises the bivariate correlations among demographics, subjective appraisals, and TMT-VR outcomes. A clear age gradient emerged: older adults showed larger spatial-selection errors, more mis-selections, and slightly longer completion times on the more demanding Part B, while years of education were unrelated to any performance index. Self-reported gaming skill was likewise unconnected to speed and improved accuracy only in the younger cohort, suggesting that controller familiarity confers little benefit once hands-free input is available.

Subjective ratings were linked to precision but not speed. Participants who reported a more favourable user-experience (UX) tended to place the ray closer to the centre of each target, and those who judged the system highly usable or acceptable committed fewer mis-selections on Part A. Because these relationships are strictly correlational, the data do not show whether positive impressions promote better pointing or whether accurate performers simply view the system more favourably. What can be concluded is that perceptual ease and spatial precision co-occur, whereas the time needed to complete each trail appears largely independent of users' experiential evaluations.

**Table 4.** Correlations: Performance between Demographics and User Evaluation

|  | Task Time A | Task time B | Accuracy A | Accuracy B | Mistakes A | Mistakes B |
|---|---|---|---|---|---|---|
| Age | 0.228 *** | 0.250 *** | 0.642 *** | 0.596 *** | 0.514 *** | 0.249 *** |
| Education | 0.023 | -0.041 | 0.059 | 0.053 | 0.104 | -0.039 |
| GSQ | 0.051 | 0.057 | 0.097 | -0.001 | -0.087 | -0.047 |
| UX | 0.023 | 0.111 | -0.264 *** | -0.279 *** | -0.212 ** | 0.093 |
| Usability | -0.028 | -0.020 | 0.000 | -0.061 | -0.149 * | -0.097 |
| Acceptability | -0.021 | -0.059 | -0.154 * | -0.092 | -0.129 * | -0.010 |
| PC | 0.038 | -0.019 | 0.166 * | 0.154 * | 0.088 | -0.009 |
| SMART | 0.009 ** | -0.055 | 0.131 * | 0.044 | 0.072 | -0.120 |
| VR | 0.179 ** | 0.190 ** | 0.333 *** | 0.264 *** | 0.155 * | 0.078 |

Note. * $p < .05$, ** $p < .01$, *** $p < .001$, GSQ = Gaming Skills Questionnaire; UX = User Experience Questionnaire; Usability = Usability Questionnaire; Acceptability = Acceptability Questionnaire ; PC =Personal Computer Use; SMART = Smartphone Application Use; VR = VR Use

Table 5 displays how individual differences colour participants' subjective impressions of the TMT-VR. Older adults reported richer overall experience (UX) and higher acceptability, yet paradoxically judged the interface slightly less usable, suggesting that age influences affective and pragmatic appraisals in opposite directions. Formal education showed no meaningful ties to any rating. Gaming-skill scores (GSQ) correlated only with usability—habitual players found the controls easier, but did not enjoy or endorse the system more than non-gamers—whereas everyday digital fluency told a clearer story: self-reported computer and smartphone use aligned positively and consistently with UX, usability, and acceptability. Taken together, these modest but reliable associations indicate that general technology familiarity, rather than gaming expertise per se, tracks with more favourable evaluations, while age shapes how engaging and how easy the platform feels.

**Table 5.** Demographics with User Evaluation

|  | UX | Usability | Acceptability |
|---|---|---|---|
| Age | 0.182 * | -0.177 ** | 0.191 ** |
| Education | -0.125 | 0.006 | 0.129 |
| GSQ | -0.125 | 0.167 * | 0.055 |
| PC | 0.155 * | 0.217 *** | 0.243 *** |
| SMART | 0.136 * | 0.334 *** | 0.419 *** |

Note. * $p < .05$, ** $p < .01$, *** $p < .001$, GSQ = Gaming Skills Questionnaire; PC = Personal Computer Use; SMART = Smartphone Application Use

*Task Performance (Mixed-Model ANOVAs)*

All behavioural indices were analysed with a 3 (Modality: eye-tracking, head-gaze, hand-controller; within-subjects) × 2 (Age Group: young, middle-aged; between-subjects) × 2 (Gaming Level: high, low; between-subjects) mixed ANOVA. Means and dispersion by condition appear in Table 6 (Interaction Mode) and Table 7 (Gaming Level and Age Group).

**Table 6**. – Performance by Interaction Mode

|  | InteractionMode | Mean (SD) | Range |
|---|---|---|---|
| AccuracyA | Eye | 22.04 (7.22) | 15.0–32.2 |
|  | Hand | 22.86 (7.98) | 15.3–38.0 |

|  | InteractionMode | Mean (SD) | Range |
|---|---|---|---|
|  | Head | 22.48 (7.73) | 15.1–37.8 |
|  | All | 22.46 (7.62) | 15.0–38.0 |
| AccuracyB | Eye | 22.19 (7.48) | 14.9–35.0 |
|  | Hand | 22.53 (7.65) | 15.3–35.9 |
|  | Head | 22.31 (7.48) | 15.1–33.7 |
|  | All | 22.34 (7.51) | 14.9–35.9 |
| TaskTimeA | Eye | 78.90 (22.08) | 38.0–149.2 |
|  | Hand | 87.49 (20.17) | 36.3–136.8 |
|  | Head | 76.76 (19.91) | 45.9–135.1 |
|  | All | 81.05 (21.17) | 36.3–149.2 |
| TaskTimeB | Eye | 94.23 (34.13) | 46.1–288.5 |
|  | Hand | 101.89 (29.70) | 48.7–206.9 |
|  | Head | 88.66 (24.12) | 32.8–144.5 |
|  | All | 94.93 (29.97) | 32.8–288.5 |
| MistakesA | Eye | 2.06 (3.08) | 0–11 |
|  | Hand | 1.47 (2.00) | 0–8 |
|  | Head | 1.12 (1.56) | 0–7 |
|  | All | 1.55 (2.32) | 0–11 |
| MistakesB | Eye | 2.62 (3.28) | 0–12 |
|  | Hand | 1.87 (2.87) | 0–12 |
|  | Head | 1.65 (1.58) | 0–8 |
|  | All | 2.05 (2.69) | 0–12 |

*Note.* Accuracy = mean selection distance (lower = better). Mistakes = total selection errors.

**Table 7.** – Performance by Gaming Level and Age Group

|  | **Gaming Level** | **Age Group** | **Mean (SD)** | **SD** |
|---|---|---|---|---|
| AccuracyA | High | Middle | 30.06 (4.46) | 16.9–38.0 |
|  |  | Young | 15.61 (0.28) | 15.1–16.3 |
|  | Low | Middle | 28.93 (4.37) | 16.7–34.5 |
|  |  | Young | 15.60 (0.27) | 15.0–16.3 |
|  | Both | Middle | 29.49 (4.43) | 16.7–38.0 |
|  |  | Young | 15.61 (0.27) | 15.0–16.3 |
| AccuracyB | High | Middle | 29.28 (4.07) | 17.0–35.9 |
|  |  | Young | 15.57 (0.27) | 14.9–16.2 |
|  | Low | Middle | 29.29 (4.60) | 16.3–33.5 |
|  |  | Young | 15.59 (0.22) | 15.2–16.1 |

|  | **Gaming Level** | **Age Group** | **Mean (SD)** | **SD** |
|---|---|---|---|---|
|  | Both | Middle | 29.28 (4.32) | 16.3–35.9 |
|  |  | Young | 15.58 (0.25) | 14.9–16.2 |
| TaskTimeA | High | Middle | 92.03 (23.23) | 38.0–149.2 |
|  |  | Young | 73.20 (17.10) | 42.3–120.0 |
|  | Low | Middle | 83.58 (21.57) | 36.3–136.8 |
|  |  | Young | 75.81 (17.45) | 43.1–119.5 |
|  | Both | Middle | 87.81 (22.71) | 36.3–149.2 |
|  |  | Young | 74.47 (17.24) | 42.3–120.0 |
| TaskTimeB | High | Middle | 108.14 (40.59) | 32.8–288.5 |
|  |  | Young | 85.95 (22.05) | 45.6–150.9 |
|  | Low | Middle | 100.10 (23.99) | 64.7–162.3 |
|  |  | Young | 86.00 (24.12) | 48.5–144.9 |
|  | Both | Middle | 104.12 (33.43) | 32.8–288.5 |
|  |  | Young | 85.98 (22.98) | 45.6–150.9 |
| MistakesA | High | Middle | 2.70 (2.62) | 0–11 |
|  |  | Young | 0.25 (0.88) | 0–5 |
|  | Low | Middle | 2.95 (2.73) | 0–11 |
|  |  | Young | 0.37 (0.72) | 0–3 |
|  | Both | Middle | 2.83 (2.67) | 0–11 |
|  |  | Young | 0.31 (0.80) | 0–5 |
| MistakesB | High | Middle | 2.75 (3.57) | 0–12 |
|  |  | Young | 1.12 (1.43) | 0–5 |
|  | Low | Middle | 2.91 (2.87) | 0–12 |
|  |  | Young | 1.46 (2.05) | 0–10 |
|  | Both | Middle | 2.83 (3.22) | 0–12 |
|  |  | Young | 1.28 (1.76) | 0–10 |

Note. "Both" = high + low gamers combined within each age group; lower Accuracy values indicate greater spatial precision.

Accuracy in Task A

The mixed-model ANOVA showed that pointing precision differed significantly by Modality, $F(2, 219) = 13.94$, $p < .001$, $\eta^2_p = .11$, and even more so by Age Group, $F(1, 219) = 460.57$, $p < .001$, $\eta^2_p = .68$.1 Gaming experience was not a reliable factor, $F(1, 219) = 1.91$, $p = .17$, $\eta^2_p = .01$. None of the two-way interactions—Modality × Gaming, $F(2, 219) = 0.89$, $p = .41$, $\eta^2_p = .01$; Modality × Age, $F(2, 219) = 1.40$, $p = .25$, $\eta^2_p = .01$; Gaming × Age, $F(1, 219) = 2.30$, $p = .13$, $\eta^2_p = .01$—or the three-way interaction, $F(2, 219) = 1.74$, $p = .18$, $\eta^2_p = .02$, reached significance.

Post-hoc (Bonferroni) comparisons clarified the modality effect. Eye-tracking yielded the lowest error (mean difference vs. hand-controller = –0.48 z), $p < .001$; Cohen's $d = 1.45$, and was also more precise than head-gaze (–0.22 z), but $p = .055$ (see Figure 3). Head-gaze out-performed the controller (+0.26 z), *p* = .012. For age, middle-aged adults were dramatically less accurate than young adults (see Figure 3). Thus, while gaze-based input—especially eye-tracking—markedly enhances spatial accuracy, a substantial age advantage persists irrespective of interaction technology.

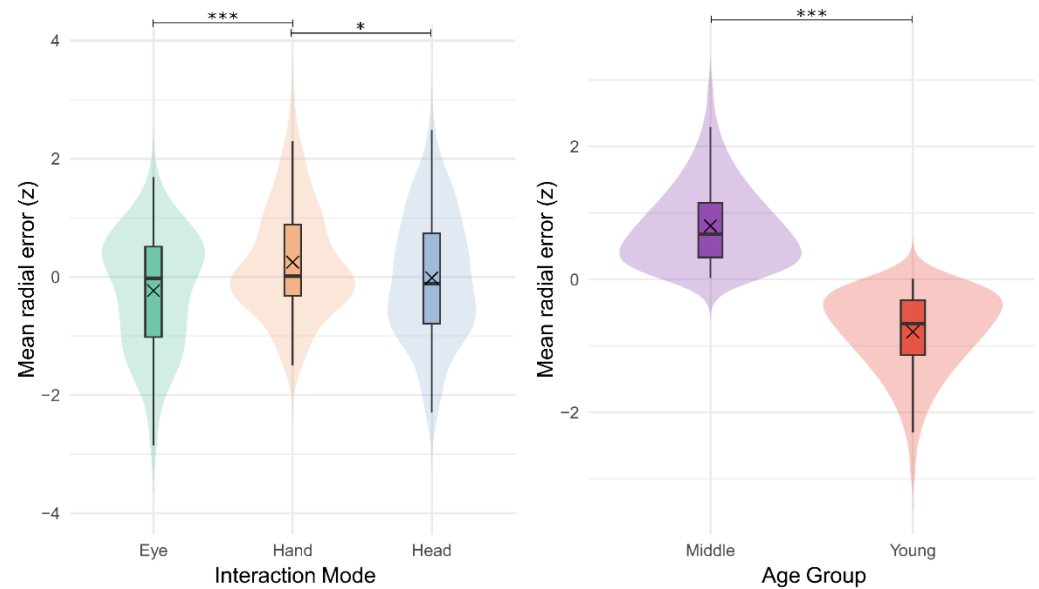

**Figure 3.** Accuracy A: spatial error (z-scores) by Interaction Mode (Left) and Age Group (Right). Only Bonferroni-significant pair-wise contrasts are shown. * p < .05, *** p < .001.

Accuracy in Task B

The mixed-model ANOVA again confirmed that spatial accuracy differed by Interaction Modality, $F(2, 219) = 4.38$, $p = .014$, $\eta^2_p = .04$, and—much more strongly—by Age Group, $F(1, 219) = 410.30$, $p < .001$, $\eta^2_p = .65$. Gaming experience exerted no reliable effect, $F(1, 219) = 0.77$, $p = .38$, $\eta^2_p = .01$, and none of the two- or three-way interactions approached significance (largest $F = 0.79$, $p = .46$, $\eta^2_p \leq .02$).

Bonferroni-adjusted comparisons showed that eye-tracking was still the most precise input: its mean error was 0.28 z-units lower than the hand-controller, *p* = .011, Cohen's *d* = 0.13 (see Figure 4). The eye-tracking versus head-gaze contrast (–0.12 z) was non-significant, *p* = .59, as was head-gaze versus controller (+0.16 z), *p* = .31. Mirroring Task A, middle-aged adults were considerably less accurate than young adults (see Figure 4). Thus, although modality differences were smaller than in the simple trail, gaze-based input—particularly eye-tracking—still afforded a measurable accuracy benefit, and the substantial age advantage persisted.

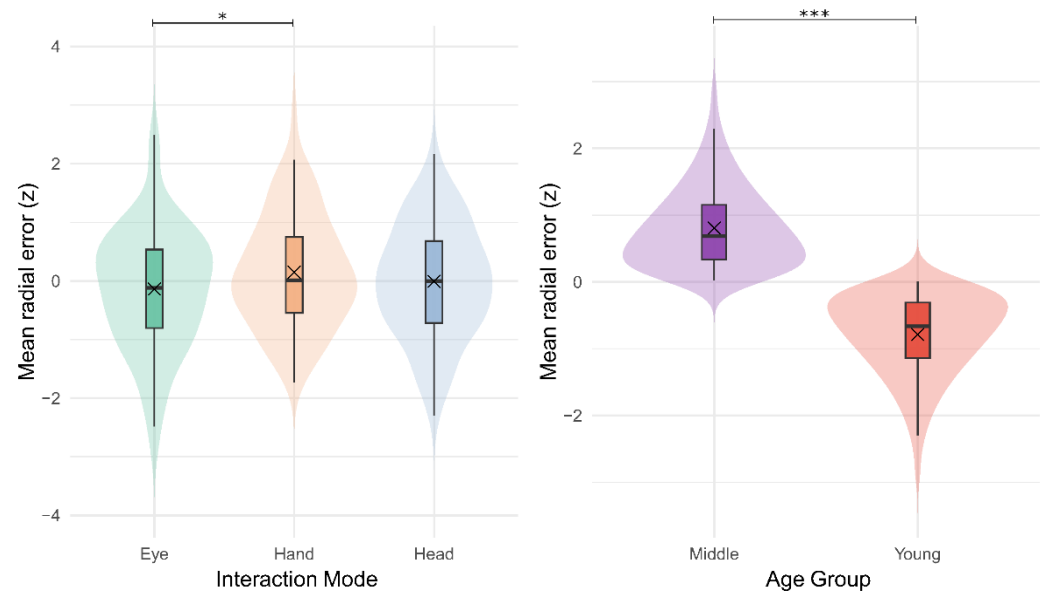

**Figure 4.** Accuracy B: spatial error (z-scores) by Interaction Mode (Left) and Age Group (Right). Only Bonferroni-significant pair-wise contrasts are shown. * p < .05, *** p < .001.

Completion Time in Task A

The mixed-model ANOVA showed that trail-completion speed varied by Interaction Modality, $F(2, 219) = 6.73$, $p = .001$, $\eta^2_p = .06$, and Age Group, $F(1, 219) = 23.22$, $p < .001$, $\eta^2_p = .10$. The Gaming-Level main effect was negligible, $F(1, 219) = 0.75$, $p = .39$, $\eta^2_p = .01$, and the critical Modality × Age interaction was not reliable, $F(2, 219) = 0.01$, $p = .99$, $\eta^2_p < .001$. The only interaction that reached significance was Gaming Level × Age Group, $F(1, 219) = 4.12$, $p = .044$, $\eta^2_p = .02$; all other interaction terms were nonsignificant (largest $F = 1.32$, $p = .27$).

Bonferroni-adjusted pairwise contrasts clarified the modality main effect. Eye-tracking cut ~8.6 s off the completion time relative to the hand-controller, $p = .013$; Cohen's $d = 0.43$ (see Figure 5). The head-gaze condition was reliably faster than the controller as well, $p = .002$; $d = 0.49$. Eye-tracking and head-gaze did not differ, $p = 1.00$. The age main effect reflected a sizeable advantage for young adults over middle-aged adults (see Figure 5). Follow-ups on the Age × Gaming interaction showed that the speed gap was driven chiefly by high-gaming participants: middle/high gamers were slower than young/high gamers, $p < .001$, $d = 0.42$, whereas no reliable difference emerged between the two low-gaming groups ($p = .32$) (see Figure 6).

In sum, hands-free gaze control reduced Trail A completion times relative to conventional controllers, and did so across both age cohorts. Nevertheless, young adults retained a clear overall speed advantage, which was most pronounced among highly experienced gamers.

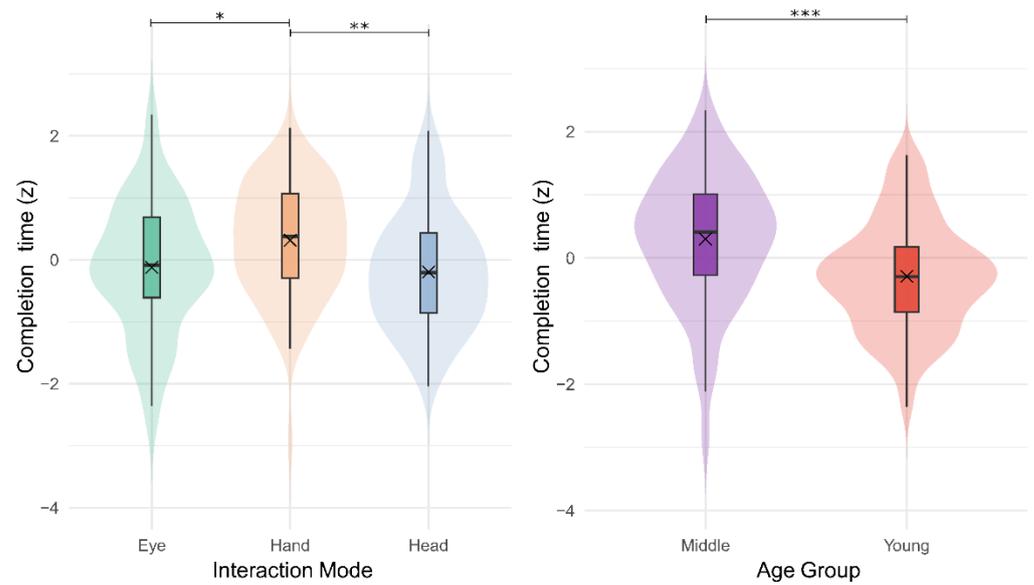

**Figure 5.** Completion Time A: task-time (z-scores) by Interaction Mode (Left) and Age Group (Right). Only Bonferroni-significant pair-wise contrasts are shown. * p < .05, ** p < .01, *** p < .001.

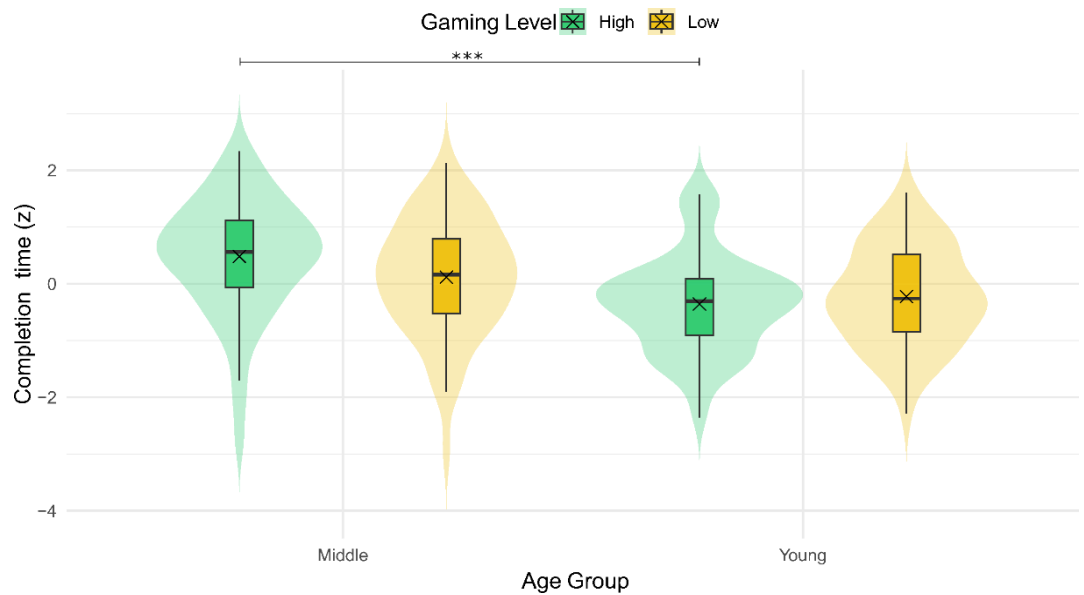

**Figure 6.** Completion Time A: task-time (z-scores) by Age Group × Gaming Level. Only Bonferroni-significant pair-wise contrasts are shown. *** p < .001.

Completion Time in Task B

The alternating trail (Task B) again showed clear performance differences in the mixed-model ANOVA. Interaction Modality produced a medium main effect, $F(2, 219) = 4.99$, $p = .008$, $\eta^2_p = .04$, while Age Group yielded a slightly larger effect, $F(1, 219) = 25.04$, $p < .001$, $\eta^2_p = .10$. Neither Gaming Level ($F(1, 219) = 0.60$, $p = .441$, $\eta^2_p = .00$) nor any two- or three-way interactions approached significance (largest $F = 0.55$, $p = .58$, $\eta^2_p \leq .01$).

Bonferroni-corrected contrasts clarified the modality effect. Head-gaze was significantly faster than the hand-controller, $p = .007$; Cohen's $d = 0.49$ (see Figure 7). Eye-tracking was numerically quicker than the controller (Δ = 7.7 s) but did not survive the Bonferroni adjustment, $p = .12$; $d = 0.24$, nor did it differ from head-gaze ($p = .92$; $d = 0.23$).

The age effect replicated the pattern from Task A, where middle-aged adults were markedly slower than young adults (see Figure 7).

In summary, for the more demanding alternating trail, head-gaze offered the clearest speed advantage over hand control, whereas eye-tracking showed a modest, non-significant gain. Middle-aged participants consistently required more time than young adults, indicating that age-related slowing persists even with gaze-based interaction, though all groups benefited to some extent from hands-free modalities.

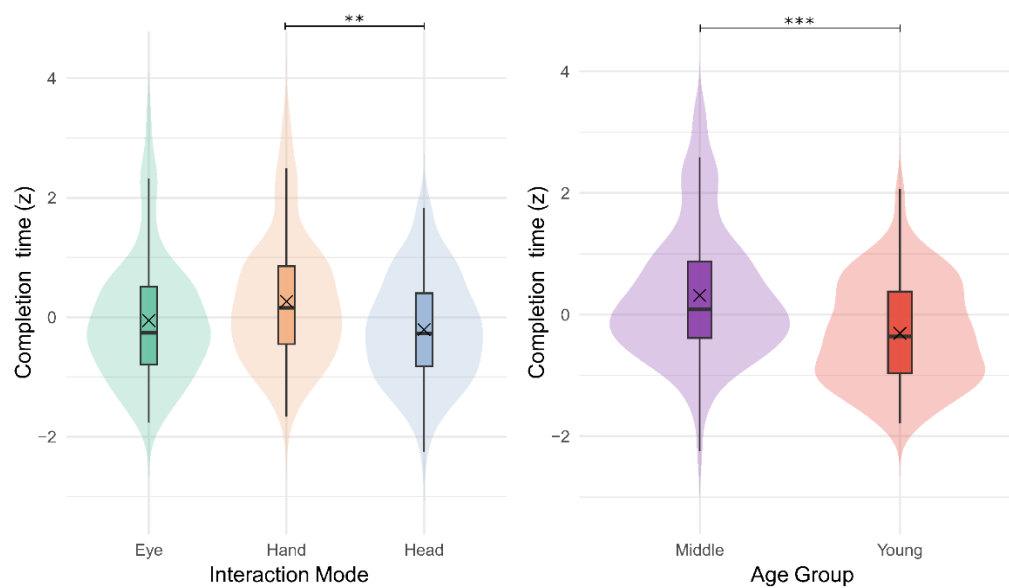

**Figure 7.** Completion Time B: task-time (z-scores) by Interaction Mode (Left) and Age Group (Right). Only Bonferroni-significant pair-wise contrasts are shown. ** $p < .01$, *** $p < .001$.

Mistakes in Task A

The mixed-model ANOVA on the number of selection errors revealed significant main effects of Interaction Modality, $F(2, 219) = 3.98$, $p = .020$, $\eta^2_p = .04$, and Age Group, $F(1, 219) = 109.62$, $p < .001$, $\eta^2_p = .33$. Gaming Level was not a significant factor, $F(1, 219) = 0.78$, $p = .38$, $\eta^2_p = .00$, and neither two- nor three-way interactions reached significance (largest $F = 2.17$, $p = .12$, $\eta^2_p \leq .02$).

Bonferroni-adjusted comparisons showed that eye-tracking produced more errors than head-gaze, $p = .016$; Cohen's $d = 0.34$ (see Figure 8). All other modality contrasts were non-significant (eye vs. hand, $p = .36$; hand vs. head, $p = .63$). The age effect was pronounced, where middle-aged adults committed reliably more errors than young adults (see Figure 8). Because no interactions with Age or Gaming Level were significant, the modest but significant modality effect (advantage for head-gaze) and the substantial age-related difference in error frequency were consistent across gaming-experience groups.

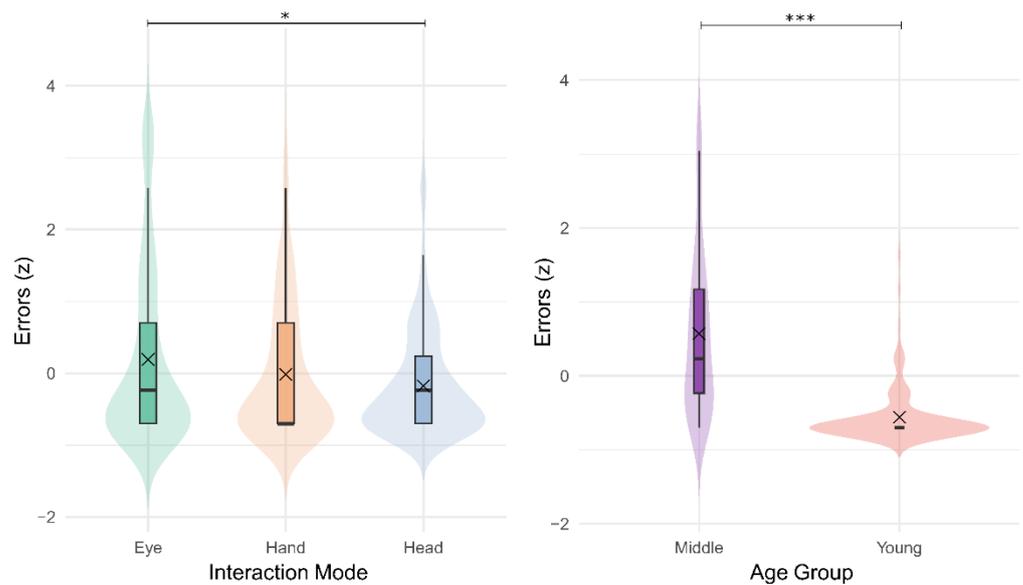

**Figure 8.** Mistakes A: error count (z-scores) by Interaction Mode (Left) and Age Group (Right). Only Bonferroni-significant pair-wise contrasts are shown. * p < .05, *** p < .001.

Mistakes in Task B

The mixed-model ANOVA confirmed that error counts differed by *Interaction Modality* and *Age Group*. Modality was significant, $F(2, 219) = 3.23$, $p = .041$, $\eta^2_p = .03$, and Age yielded a larger effect, $F(1, 219) = 22.13$, $p < .001$, $\eta^2_p = .09$. Gaming Level showed no main effect, $F(1, 219) = 1.13$, $p = .289$, $\eta^2_p = .01$. The Modality × Gaming interaction reached significance, $F(2, 219) = 3.86$, $p = .023$, $\eta^2_p = .03$, whereas Modality × Age and the three-way term remained nonsignificant (largest $F = 2.91$, $p = .057$, $\eta^2_p \leq .03$).

Bonferroni-corrected pairwise tests showed that eye-tracking produced more errors than head-gaze $p = .048$, Cohen's $d = 0.38$; the eye- vs. hand-controller contrast did not reach significance ($p = .139$), and hand vs. head was wholly nonsignificant ($p = 1.00$) (See Figure 9). Mean error rates were low overall (= 1–3 mistakes per trial). Age comparisons indicated that middle-aged adults made reliably more errors than young (See Figure 9).

Decomposing the Modality × Gaming interaction revealed that the disadvantage of eye-tracking was concentrated in low-gaming participants. Specifically, low-skill users made significantly more errors with eye-tracking than with the hand controller, $p = .014$, $d = 0.77$, and more than high-skill users operating head-gaze, $p = .016$, $d = 0.76$ (See Figure 10). Among high-gaming participants, no modality contrasts survived correction (all adjusted $p$s ≥ .84).

Taken together, the alternating-set task elicited slightly higher error rates with eye-tracking—especially for participants lacking gaming experience—and a robust age effect favouring the younger cohort, while absolute error counts remained modest across all conditions.

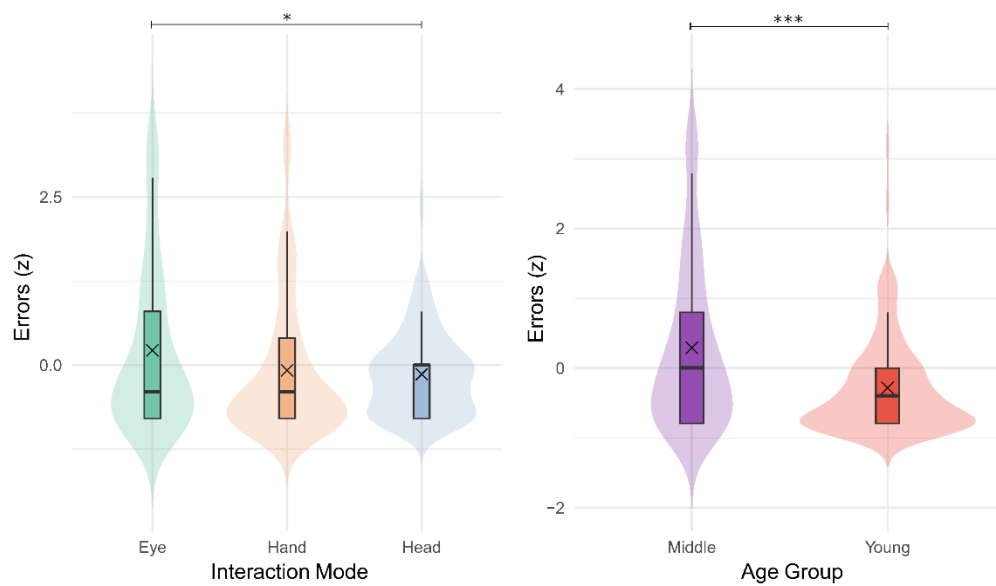

**Figure 9.** Mistakes B: error count (z-scores) by Interaction Mode (Left) and Age Group (Right). Only Bonferroni-significant pair-wise contrasts are shown. * p < .05, *** p < .001.

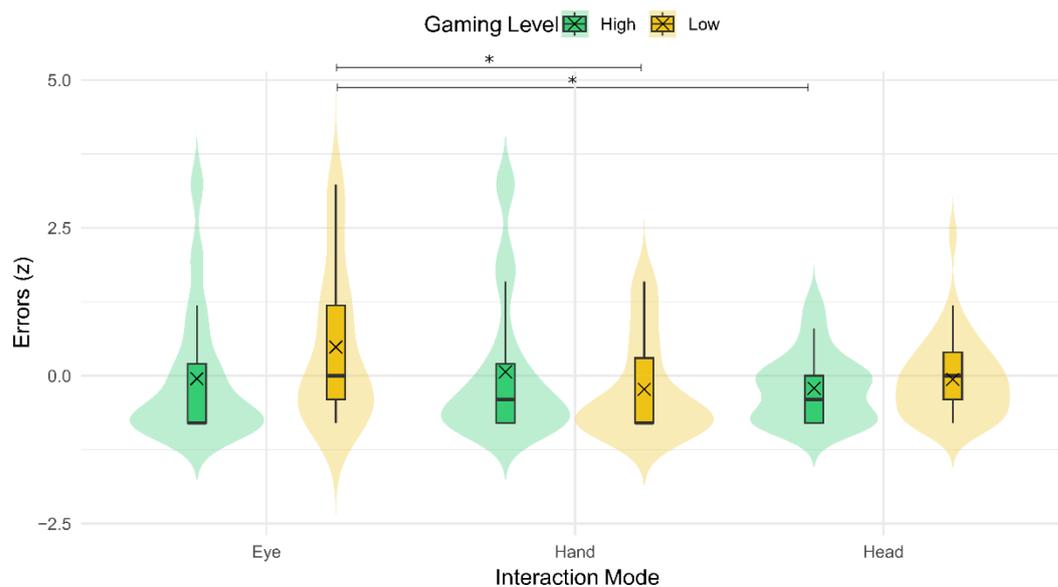

**Figure 10.** Mistakes B: error count (z-scores) by Interaction Mode × Gaming Level. Only Bonferroni-significant pair-wise contrasts are shown. * p < .05.

*Acceptability, Usability, and User Experience*

For acceptability, the mixed-model ANOVA revealed a marginal, yet non-significant Age, effect, $F(1, 227) = 3.84$, $p = .051$, $\eta^2_p = .02$, and a non-significant Gaming main effect, $F(1, 227) = 2.71$, $p = .101$, $\eta^2_p = .01$. The Age × Gaming interaction also failed to reach significance, $F(1, 227) = 0.30$, $p = .583$, $\eta^2_p = .001$. Thus, acceptability was rated similarly high by the whole sample regardless of participants' age and gaming experience.

On the contrary, perceived usability (SUS) depended on the joint influence of age and gaming expertise. The Age × Gaming interaction reached significance, $F(1, 227) = 8.54$, $p = .004$, $\eta^2_p = .04$, whereas the main effects of Age, $F(1, 227) = 3.33$, $p = .069$, $\eta^2_p = .01$, and Gaming Level, $F(1, 227) = 1.33$, $p = .249$, $\eta^2_p = .006$, were non-significant. Follow-up contrasts revealed that middle-aged participants with low gaming experience reported lower usability than their young, low-gaming counterparts, $p = .006$, $d = 0.63$; and middle-aged high-gamers $p$ = .025, $d$ = 0.54 (see Figure 11). No other pairwise comparisons survived the Bonferroni adjustment (all adjusted $ps \geq .19$), suggesting that strong gaming skills largely neutralised age-related differences in SUS ratings.

Finally, user-experience (UEQ-S) scores displayed a robust age advantage that was independent of gaming history. Middle-aged adults evaluated the session as more engaging than young adults, $F(1, 227) = 24.42$, $p < .001$, $\eta^2_p = .10$ (see Figure 11). Gaming Level, $F(1, 227) = 1.28$, $p = .259$, $\eta^2_p = .006$, and the Age × Gaming interaction, $F(1, 227) = 0.37$, $p = .544$, $\eta^2_p = .002$, were both non-significant. Taken together, these questionnaire findings indicate that the TMT-VR was viewed as highly acceptable, usable, and enjoyable across the sample, with the most favourable impressions emerging from middle-aged participants.

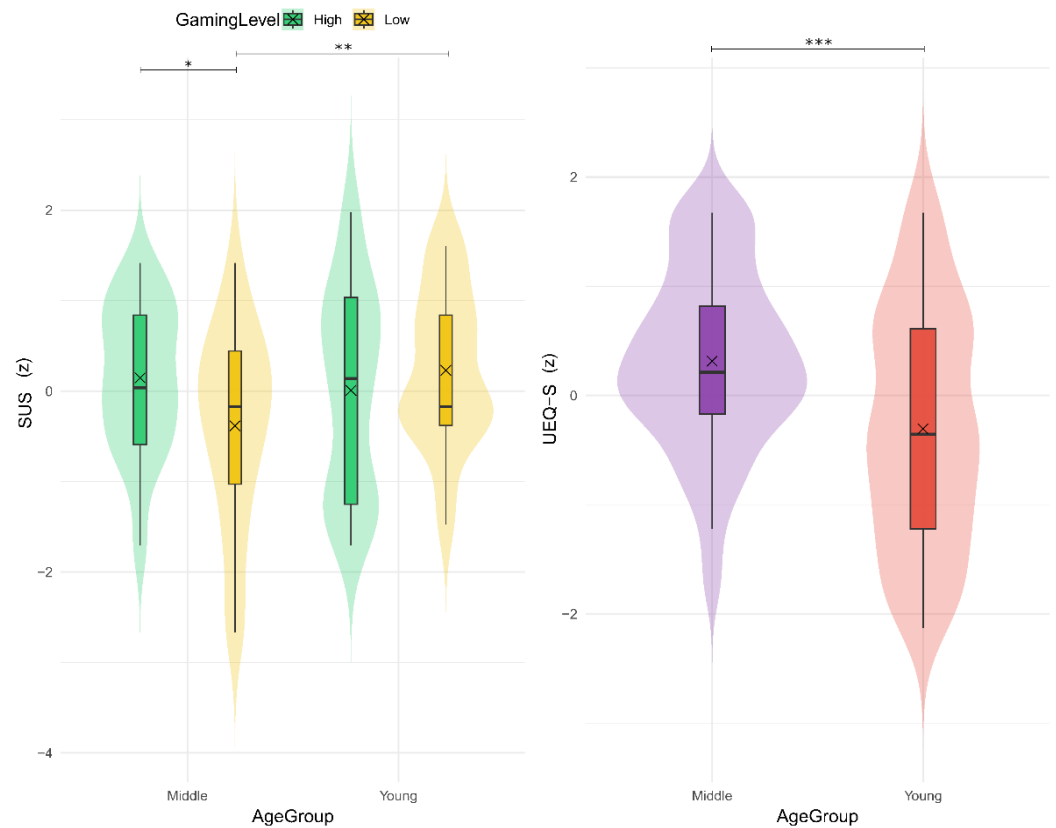

**Figure 11.** Perceived Usability and User Experience: SUS scores (z-scores) by Age Group × Gaming Level (Left) and UEQ-S scores (z-scores) by Age Group (Right). Only Bonferroni-significant pairwise contrasts are shown. * p < .05, ** p < .01, *** p < .001.

## 4. Discussion

This study investigated how three hands-free VR pointing modes—eye-tracking, head-gaze, and a six-degree-of-freedom hand controller—interact with age and gaming history to influence performance on a VR adaptation of the Trail-Making Test (TMT-VR) and to shape users' subjective evaluations of the system. Four patterns frame the ensuing discussion. First, age dominated behaviour: younger adults were consistently faster, more precise, and less error-prone than middle-aged adults. Second, eye-tracking offered the clearest technological dividend, sharpening spatial accuracy and shortening completion time on the simpler trail relative to manual control. Third, head-gaze emerged as the speed leader on the cognitively demanding alternating trail, implying that its slower, stabilised movement profile may lighten cognitive load when switching demands peak.

Fourth, user impressions were uniformly positive: participants across the spectrum deemed the assessment acceptable, engaging, and easy to handle, with only a mild usability dip in middle-aged, low-gaming individuals—a gap that vanished among their high-gaming peers. The remainder of the Discussion unpacks these themes, relating them to theories of cognitive ageing, gaze-based interaction, and VR assessment design, and concludes with practical recommendations for researchers and clinicians.

*Interaction-Modality Effects on Task Performance*

Across both trails the choice of input device shaped behaviour in distinct yet complementary ways. In the simple, sequential trail (Task A) eye-tracking produced the tightest pointing dispersion, while head-gaze also improved precision relative to the hand controller. On the cognitively heavier alternating trail (Task B) eye-tracking still secured the lowest mean error overall, but its accuracy lead over head-gaze was small and not statistically reliable—underscoring that both gaze-based methods outperform manual pointing when calibration succeeds [33,38,72]. Controllers lagged in precision on both tasks, echoing ergonomic rankings that place eye < head < hand for effort and fine spatial control [39].

Speed painted a complementary picture. On the easy trail eye- and head-gaze were equally quick and each out-paced the controller, consistent with work showing that eliminating arm transport shortens the action-selection cycle [73]. When continuous set-switching was required (Task B) head-gaze became the fastest option. A plausible explanation is that the slight inertia of head pivots promotes smoother sweeping when the ray must traverse the display repeatedly, whereas microsaccadic jitter can add re-calibration overhead to eye-tracking under high-frequency scans [43]. The controller again remained slowest, likely because holding a six-degrees-of-freedom device at chest height taxes proximal musculature and elongates movement planning [37].

Error patterns broadly paralleled the accuracy data, with head-gaze displaying a modest advantage. Eye-tracking incurred extra slips only among low-gaming participants on the complex trail—a subgroup that may lack the oculomotor discipline cultivated by fast-paced gaming and therefore overshoot during rapid alternations, as previously noted for novice users [39]. Nonetheless, absolute error counts remained low, confirming that the ray-casting design is forgiving across skill levels.

Taken together, the modality findings highlight two mechanisms that designers can exploit. Attentional alignment: coupling the selection channel to the visual focus collapses search and select into a single act, maximising precision. Ergonomic economy: when frequent directional updates are needed, the low-amplitude, momentum-buffered movements of the head can outrun both eye and hand for raw speed while avoiding the fatigue associated with prolonged arm elevation [74]. In practice, eye-tracking should be the first choice for accuracy-critical assessments when a brief, reliable calibration is feasible; head-gaze offers a robust, calibration-free alternative that excels under sustained switching demands; handheld controllers are best reserved for onboarding or contingency use when gaze-based input is not viable.

*Participant Factors: Age and Gaming Background*

Age emerged as the single most powerful predictor of performance. Across both accuracy measures the effect was enormous, and sizeable age gaps re-appeared for completion time and error count. These patterns echo classic processing-speed and motor-slowing accounts of midlife cognition [75], but they also highlight that even when gross motor demands are removed, older adults still accumulate more cumulative pointing noise than younger peers. The implication for assessment design is straightforward: VR adaptations do not erase age norms; they change the scale, not the rank order.

Gaming experience, by contrast, showed only two qualified interactions. First, in Task-time A an Age × Gaming crossover revealed that the speed advantage of youth was driven almost entirely by high-gaming participants—young/high gamers were the fastest group, whereas middle/high gamers were no quicker than their low-gaming age-mates. One plausible interpretation is skill-transfer asymmetry: years of fast-paced gaming may hone visuomotor timing in early adulthood but confer diminishing returns once natural motor speed begins to plateau in midlife [76].

Second, in Mistakes B a Modality × Gaming interaction showed that low-gaming participants committed extra errors when using eye-tracking, whereas high-gaming players distributed their errors evenly across inputs. This finding dovetails with reports that novices need longer to stabilise oculomotor control in gaze-based interfaces [39]. Crucially, the effect was confined to the demanding alternating trail, suggesting that limited gaming experience becomes a liability only when cognitive load and sensorimotor novelty coincide.

Together these results refine two literatures. From a cognitive-aging perspective they confirm that large age differences persist in immersive, hands-free settings, underscoring the need for age-stratified norms even in next-generation tests. From a skill-transfer standpoint they show that habitual gaming *can* sharpen speed and dampen oculomotor error—but only for the young or when interface demands are high. For clinical VR assessments this means that simple tutorial exposure is sufficient for most users, yet optional extended practice may be warranted for low-gaming, older clients if gaze interaction is required.

*Subjective Appraisals and Their Link to Objective Performance*

Participants evaluated the TMT-VR very positively. Acceptability ratings were uniformly high, echoing evidence that modern head-mounted displays now routinely clear the comfort and cybersickness bar for adult users [10,14]. Perceived usability, however, displayed a subtle Age × Gaming interaction. Among highly skilled gamers, young and middle-aged adults rated the interface equally easy, but within the low-gaming subgroup middle-aged adults judged it less usable than their younger counterparts. A likely explanation is *expectation mismatch*: young non-gamers often approach VR with novelty enthusiasm, whereas their older peers arrive with concrete efficiency standards shaped by decades of workplace software use [77]. When those standards are not instantly met—e.g., during eye-tracking calibration—usability scores dip.

User-experience (UEQ-S) told a different story: middle-aged adults consistently reported richer engagement than the young, a pattern that mirrors age-related preferences for immersive media that feel purposeful rather than merely entertaining [49,78]. In other words, the very group that moved more slowly and less precisely found the session *more* absorbing. This divergence between affective appraisal and motor output cautions against treating "better experience" as a proxy for "better performance."

Bivariate correlations reinforced that point. Higher UX, usability, and acceptability scores were modestly associated with finer spatial precision and fewer errors on the simple trail, yet they were essentially unrelated to completion time. The dissociation suggests two quasi-independent pathways: (i) an embodied-fluency route, where an interface that *feels* smooth encourages careful but confident aiming, and (ii) a processing-speed route that is largely immune to subjective comfort. For designers, this means polishing micro-interactions (cursor stability, feedback latency) can yield measurable accuracy gains without necessarily accelerating task throughput; conversely, slimming task duration will not automatically boost user delight.

In sum, the questionnaire data complement the performance findings rather than mirroring them. They show that middle age is not a barrier to enjoyment, that gaming practice buffers usability complaints, and that experiential quality aligns better with *how well* users hit the targets than with *how fast* they do so.

*Design and Clinical Implications*

Eye-tracking delivered the finest spatial precision on both trails, but it did not dominate every metric. Head-gaze equalled eye-tracking on speed in the easy trail, surpassed it on the harder alternating trail, and produced the fewest selection errors overall. Accordingly, the two hands-free modalities should be treated as complementary rather than hierarchical. When the hardware supports a quick, single-shot calibration, eye-tracking remains the method of choice for accuracy-critical applications—its tight coupling between gaze and attentional focus minimises mean selection error [38]. Yet calibration can fail in the very populations most in need of accessible assessment: people sitting under uneven lighting (for screen based eye-trackers), wearing progressive lenses (on glasses or contact) and/or heavy eyeliner (for VR). In those situations head-gaze becomes the default, not merely a fallback. Because it is calibration-free, it can be launched within seconds; because it demands only slow, stable neck movements, it preserved accuracy on the simple trail and emerged as the fastest option when cognitive switching demands were highest.

Six-degree-of-freedom hand controllers still have a place, but that place is narrow. They are useful during the onboarding tutorial, where familiar haptic triggers quell first-contact anxiety, and they serve as an emergency option for clients with severe cervical immobility or ocular conditions that preclude stable gaze input. Critically, a brief probe at the beginning of each session can identify which modality (eye or head) is technically viable; once that check is complete, there is no need to expose users to all three inputs. Locking in a single, ergonomically appropriate method reduces cognitive overhead and keeps session time within clinical constraints.

Raw scores, however, cannot be interpreted in isolation. The very large age effects observed across accuracy, speed, and error rates echo the paper-and-pencil TMT and make age-stratified norms essential [1]. Because gaming experience interacted with age on completion time (easy trail) and error rate (alternating trail), examiners should also collect a quick digital-fluency marker—weekly game hours or the brief Gaming Skills Questionnaire—so that future scoring software can place each patient in the correct joint reference band (age × gaming fluency) as larger normative datasets accumulate.

Three practical refinements flow directly from the data. First, although a five-minute guided tutorial sufficed for most users, low-skill participants still made extra errors with eye-tracking on the alternating trail. An adaptive tutorial that automatically re-runs calibration and two sample selections whenever dwell time on a target exceeds one second should close that gap without lengthening the visit. Second, capping each trail at roughly two minutes—and the full cognitive block at about fifteen—limits the neck and shoulder fatigue that can build up during prolonged head fixation or continuous controller grip [79,80]. Third, inserting a brief standing stretch between trails and flashing a subtle "blink now" cue on loading screens helps preserve oculomotor precision, especially in older adults, and was well tolerated in earlier work [14].

In short, begin with a rapid viability check: if eye-tracking calibrates cleanly, use it for its accuracy edge; if not, switch to head-gaze, which offers strong speed and error-control benefits with zero calibration cost. Reserve controllers for onboarding and rare contingency cases. Log age and gaming-fluency metadata, keep sessions short and ergonomically paced, and include micro-breaks that refresh both neck posture and oculomotor stability. Following this workflow allows clinicians and HCI practitioners to deploy the TMT-VR with minimal friction while maintaining measurement fidelity across a broad adult population.

*Limitations and Future Studies*

The present findings should be interpreted in light of several constraints. First, the sample stopped short of the older-adult range that the Trail Making Test is most often used to screen; none of the volunteers were over 60 years of age, and all were free of neurological or psychiatric diagnoses. Whether the pronounced age effects reported here scale linearly into late adulthood therefore remains an open question. Second, performance was indexed purely behaviourally. Although gaze vectors were captured for the two hands-free modalities, we did not log raw ocular kinematics, pupil dynamics, or other physiological markers that could illuminate the micro-mechanisms behind the observed speed-accuracy trade-offs. Third, the study was conducted in a controlled laboratory setting with fixed lighting and noise levels. Real-world deployments—in clinics, rehabilitation wards, or patients' homes—will inevitably introduce distractions that could magnify or attenuate modality differences.

These gaps chart a clear agenda for subsequent work. An obvious next step is to extend data collection to older and clinical cohorts, beginning with adults over 65 years and progressing to groups with mild cognitive impairment or early neurodegenerative disease. At the technological level, logging continuous eye-movement traces would allow researchers to disentangle calibration drift, ocular tremor, and strategic scanning patterns, and to explore adaptive difficulty schemes that increase dwell thresholds or target spacing in real time as oculomotor noise rises. Longitudinal designs are also warranted: repeated administrations over months would reveal whether the modest learning effects seen during onboarding plateau, compound, or interact with disease progression. Together, these extensions will determine how far the present ergonomic insights generalise and refine the TMT-VR into a mature, clinically deployable tool.

## 5. Conclusions

The present study demonstrates that the multimodal TMT-VR delivers a valid, well-tolerated measure of executive functioning across gaze-based and manual input modes. Eye-tracking furnished the greatest spatial precision and matched head-gaze for speed on the straightforward Trail A, whereas head-gaze moved ahead on the cognitively taxing Trail B and recorded the fastest speed and the fewest selection errors overall. Hand-held controllers were consistently slower and less accurate than either gaze-driven alternative. Design-wise, eye-tracking should be preferred when pinpoint spatial precision is the priority, but head-gaze should be opted as the inclusive, calibration-free option that preserves accuracy and enables faster completion time under heavier cognitive load. Age remained the dominant determinant of performance: younger adults out-paced and out-performed middle-aged adults on every behavioural index (spatial accuracy, speed, and errors), confirming the necessity of age-stratified norms in VR just as in paper-and-pencil testing. Usability, user-experience, and acceptability ratings were high throughout, with only a modest usability dip for middle-aged, low-gaming individuals. Taken together, these findings position the TMT-VR—driven by superior hands-free inputs—as an accurate, engaging, and ergonomically adaptable assessment whose scores must still be interpreted against age-specific, and ideally gaming-fluency–stratified, reference data.

original draft preparation, E.G., P.V., K.G. and P.K.; writing—review and editing, E.G., P.V., K.G., E.O., C.N. and P.K.; visualization, P.K.; supervision, P.K.; project administration, E.O., C.N. and P.K.; funding acquisition, E.O., C.N. and P.K. All authors have read and agreed to the published version of the manuscript.

**Funding:** This research received no external funding.

**Institutional Review Board Statement:** This study was conducted in accordance with the Declaration of Helsinki and approved by the Ad-hoc Ethics Committee of the Psychology Department of the American College of Greece (KG/0224, 28 February 2024).

**Informed Consent Statement:** Informed consent was obtained from all subjects involved in the study.

**Data Availability Statement:** The data presented in this study are available on request from the corresponding author. The data are not publicly available due to ethical approval requirements.

**Acknowledgments:** This study received financial support by the ACG 150 Annual Fund. Panagiotis Kourtesis designed and developed the TMT-VR.

**Conflicts of Interest:** The authors declare no conflicts of interest.